\documentclass[aps,prc,floatfix,twocolumn,showpacs,superscriptaddress]{revtex4}

\usepackage{graphicx,amsmath,amssymb,bm,nicefrac}

\begin{document}

\title{Local and global effects of beta decays on  r-process}

\author{O.\ L.\ Caballero}

\affiliation{ExtreMe Matter Institute EMMI, GSI Helmholtzzentrum f{\"ur} Schwerionenforschung GmbH, 64291 Darmstadt, Germany}

\affiliation{Institut f{\"u}r Kernphysik,
 Technische Universit{\"a}t Darmstadt, 64289 Darmstadt, Germany}

\affiliation{Department of Physics, University of Guelph, Guelph, Ontario N1G 2W1, Canada}

\author{A.\ Arcones}

\affiliation{Institut f{\"u}r Kernphysik, Technische Universit{\"a}t Darmstadt, 64289 Darmstadt, Germany}

\affiliation{GSI Helmholtzzentrum f\"ur Schwerionenforschung, Planckstr. 1, 64291 Darmstadt, Germany}

\author{I. N. Borzov}

\affiliation{Joint Institute for Nuclear Research, 141980 Dubna, Russia}

\author{K. Langanke}

\affiliation{GSI Helmholtzzentrum f\"ur Schwerionenforschung,
 Planckstr. 1, 64291 Darmstadt, Germany}

\affiliation{Institut f{\"u}r Kernphysik,
 Technische Universit{\"a}t Darmstadt, 64289 Darmstadt, Germany}

\affiliation{Frankfurt Institute for Advanced Studies,
 64289 Frankfurt, Germany}

\author{G.\ Mart{\'i}nez-Pinedo }

\affiliation{Institut f{\"u}r Kernphysik,
 Technische Universit{\"a}t Darmstadt, 64289 Darmstadt, Germany}

\affiliation{GSI Helmholtzzentrum f\"ur Schwerionenforschung,
Planckstr. 1, 64291 Darmstadt, Germany}

\date{\today} 

\begin{abstract}
Nuclear beta decay rates are an essential ingredient in simulations of the astrophysical r-process. Most of these rates still rely on theoretical modeling. However, modern radioactive ion-beam facilities have allowed to measure beta half lives of some nuclei on or close to the r-process path. These data indicate that r-process half lives are in general shorter than anticipated in the standard theoretical predictions based on the Finite Range Droplet Model (FRDM). The data have also served as important constraints for improved predictions of half lives based on continuum QRPA calculations on top of the energy-density functional theory. Although these calculations are yet limited to spherical nuclei, they include the important r-process waiting point nuclei close to and at the neutron magic numbers $N=50, 82$ and 126. We have studied the impact of these new experimental and theoretical half lives on r-process nucleosynthesis within the two astrophysical sites currently favored for the r process: the neutrino-driven wind from the freshly born neutron star in a supernova explosion and the ejecta of the merger of two neutron stars. We find that the, in general, shorter beta decay rates have several important effects on the dynamics of r-process nucleosynthesis. At first, the matter flow overcomes the waiting point nuclei faster enhancing matter transport to heavier nuclei. Secondly, the shorter half lives result also in a faster consumption of neutrons resulting in important changes of the conditions at freeze-out with consequences for the final r-process abundances. Besides these global effects on the r-process dynamics, the new half lives also lead to some local changes in the abundance distributions. \end{abstract}


\maketitle

%------------------------------------------------------------------------------------------------------------------------

\section{Introduction}

\label{sec:introduction}

Although the actual astrophysical site of the r-process is still not known with certainty, it is commonly accepted that the process occurs in an explosive environment of relatively high temperatures ($T \approx 10^9$ K) and very high neutron densities ($> 10^{20}$cm$^{-3}$) \cite{Cameron,BBFH,Cowan&Thielemann}. Under such conditions, neutron captures are much faster than competing beta decays and the r-process path in the nuclear chart is expected to proceed through a chain of extremely neutron-rich nuclei with relatively low and approximately constant neutron separation energies ($S_n\lesssim$ 3 MeV). Due to the particular strong binding of nuclei with magic neutron numbers, the neutron separation energies show discontinuities at the magic numbers N=50, 82, and 126. As a consequence the r-process matterflow slows down when it reaches these neutron-magic nuclei and has to wait for several beta decays (which are also longer than for other nuclei on the r-process path) to occur until further neutron captures are possible carrying the massflow to heavier nuclei. Thus matter is accumulated at these r-process waiting points associated with the neutron numbers N=50, 82, and 126 leading to the wellknown peaks in the observed r-process elemental abundance distribution. 

Due to their extreme neutron excesss, most nuclei on the r-process path have not yet been produced in the laboratory and hence their properties are experimentally unknown and must be theoretically estimated for r-process simulations. The relevant nuclear input to such studies are masses, which determine the r-process path in the nuclear chart, and beta halflives, which fixes the time the r-process massflow needs to transmute seed nuclei to heavy nuclei including the transactinides. Neutron capture and photodissociation rates are essential, once the r-process, in the classical picture, drops out of ($n,\gamma$)-($\gamma,n$) equilibrium or, if such equilibrium is not achieved, also during the entire nucleosynthesis process. Despite their importance only a few halflives of r-process nuclei are experimentally known. Exploiting opportunities at radioactive ion-beam facilities experimentalists have recently been able to measure the halflives of some key r-process nuclei at or in the vicinity of the magic neutron numbers $N=50$ \cite{Hosmer2005,Madurga.exp:2012,Quinn.exp:2012,Hosmer:2010} and $N=82$ \cite{Kratz86,Kratz98,Pfeiffer01,Nishimura.exp:2012}. Although measurements of halflives of r-process nuclei at the  $N=126$ shell closure have yet been impossible, experiments at GSI have succeeeded to provide valuable information for nuclei towards the third r-process peak \cite{Benlliure12}.

The r-process halflive measurements have important consequences for both, the nuclear models and r-process simulations. At first,  the halflive data clearly pointed to shortcomings in previous theoretical model calculations. The halflives most frequently adopted in r-process simulations are based on the Quasi Random Phase Approximation (QRPA)\cite{Moeller.etal:2003,Engel:1999tr,Borzov:2000,Borzov:2003,Borzov:2008, Borzov:2011,Fang:2013uwa}. The new  halflive data led to an improved energy density functional adopted by Borzov to selfconsistently calculate halflives of spherical r-process nuclei \cite{Borzov:2003,Borzov:2008, Borzov:2011}. In particular, Borzov succeeded to obtain quite good agreement with the available experimental data. It is also important to note that the halflife predictions by Borzov agree well with those obtained by more sophisticated models which, however, like the Hartree-Fock-Bogoliubov model is restricted to spherical even-even nuclei, or the interacting shell model to r-process waiting point nuclei with magic neutron numbers \cite{MartinezPinedo:1999rx,CuencaGarcia:2008dm,Zhi:2013hg}. The recent calculations also indicate that besides the dominating Gamow-Teller transitions also forbidden transitions contribute to the halflives, which is most relevant for the nuclei in the vicinity of the $N=126$ shell closure \cite{CuencaGarcia:2008dm,Suzuki_theo:2012}. 

Importantly, the experimental data are on average significantly shorter than predicted by the global models. Several r-process simulations have shown that replacing half-lives predicted by global models by experimental values (or by improved theoretical values in the vicinity of the experimentally studied nuclei e.g. \cite{Niu.theo:2013}) can have interesting consequences for the process dynamics as well as the final abundance distributions. In fact, these simulations performed assuming different r-process sites (neutrino-driven wind model, neutron star merger) showed that the shorter halflives result in faster mass flows towards heavier nuclei \cite{Nishimura.exp:2012,Hosmer:2010,Madurga.exp:2012}. On the other hand, Surman et al \cite{Surman.beta:2012} studied the impact of beta decay half-lives on the r-process with a different approach. They identified the nuclei that have the largest impact on the final abundances by performing individual changes in the half-lives by a factor of 10 up and down. Their study confirmed that the nuclei with larger impact in the production of heavy elements are the ones around the waiting points $N=82$ and $N=126$, and those along the r-process path.

The aim of the present paper is to study the impact of the new generally shorter half lives on the r-process dynamics and abundance distribution. In our simulations we have adopted experimental half lives whenever available. These have been supplemented by two different sets of theoretical half lives for those nuclei for which no experimental data yet exist. Besides studying the effects that different theoretical beta decay rates have locally on the production of nuclei in the region where the rates are changed, we also investigate global effects which can be introduced by changes of the astrophysical conditions at which the r-process operates for different half lives. In particular we correlate these effects to the evolution of astrophysical quantities like the neutron density, which has been proven to have a significant impact on  the final r-process abundances \cite{Arcones.MartinezPinedo:2011}. Our simulations have been performed for the astrophysical sites currently favored for r-process nucleosynthesis. First, we adopt trajectories which correspond to the neutrino-driven wind model within a supernova explosion and reflect situations of a 'cold r-process' (i.e. no $(n,\gamma) - (\gamma,n)$ equilibrium is reached), and 'hot r-process' (this corresponds to the classical picture where the equilibrium is achieved). Second, we study the r process for trajectories derived from a hydrodynamical simulation of neutron star mergers. This study corresponds to a cold r-process with a neutron-to-seed ratio, which is significantly larger than for the case of the neutrino-driven wind scenario and allows for fission to play an important role for the final abundance distributions.

Our nuclear input is based on the global FRDM model of M\"oller and collaborators \cite{Moeller.etal:2003}. In our reference calculations we used this input, however, replacing beta decay rates by experimental values whenever possible. In a second set of calculations we have replaced the FRDM half lives by those obtained within the energy density functional model of Borzov covering a large range of nuclei in the mass range around the magic proton numbers $Z=28, 50, 82$ \cite{Borzov:2003}, the magic neutron number $N=126$ \cite{Borzov:2011}, and for neutron-rich nuclei with $Z=40-50, N <82$ \cite{Borzov:2008} and $Z=60-70$ \cite{Borzovnew}. We stress that these calculations give good agreement with all new experimental halflives relevant to r-process nucleosynthesis. Furthermore by using the same parametrizations of the density functional, as constrained by the data, we expect also a larger predictive power for those neutron-rich nuclei, which were not yet accessible experimentally, than obtained in previous halflife compilations. As a drawback, the energy functional calculations of Borzov assumed sphericity of the nuclei. Hence the model can be applied to a large set of nuclei, including those in the vicinity of the magic neutron numbers which as waiting points have a strong impact on the r-rocess nucleosynthesis, but not to deformed nuclei. This explains why we have used this model for a large portion, but not the entire nuclear chart relevant for r-process simulations.

Our manuscript is organized as follows. In the next section we describe in some detail the astrophysical scenarios we assumed for our r-process studies as well as the nuclear reaction network and the nuclear physics input, where particular emphasis is put on the adopted set of beta halflives. Section III discusses the effect of the improved halflife set on the r-process abundance patterns. We discuss in details the results for the cold r-process scenario, finding both global and local changes in the abundances. We show that these findings are not restricted to the cold r-process site, but are also observed for other scenarios like the hot r-process, and neutron star mergers. In section IV we summarize our results and give an outlook for future work.

%------------------------------------------------------------------------------------------------------------------------

\section{Nuclear physics and astrophysical input}

\label{sec:input}

%------------------------------------------------------------------------------------------------------------------------

\subsection{Nuclear reaction network and beta decays}
\label{sec:network} 

In our nucleosynthesis studies we use the same reaction network and setup as in \cite{Arcones.MartinezPinedo:2011,Arcones.Bertsch.2012}. The initial abundance evolution, from nuclear statistical equilibrium ($T\approx 10-8$~GK) to charged-particle reaction freeze-out ($T\approx 3$~GK), is calculated with a complete network that includes over 3000 nuclei and the reactions among them mediated by the strong and electromagnetic interaction. Furthermore we consider nuclear beta decays as well as reactions on nucleons mediated by the weak interaction (electron/positron captures and neutrino-induced reactions). At the lower temperatures during the r-process, the only relevant reactions are neutron capture, photo-dissociation, beta and alpha decay, and fission. We follow the evolution of r-process matter by a network (described in \cite{Arcones.MartinezPinedo:2011}) that includes these relevant reactions for more than 5000 nuclei. Our numerical treatment is quite efficient. In particular, it is well suited to accurately describe the r-process freeze-out which is characterized by the consumption of neutrons and the subsequent beta decay to stability. We define the r-process freeze-out as the moment in the evolution when the neutron-to-seed ratio drops to one, i.e., $Y_n/Y_{\mathrm{seed}} = 1$. In \cite{Arcones.MartinezPinedo:2011} the nuclear masses and neutron capture cross sections were consistently varied within the r-process network, while keeping the same set of beta decay rates. In this paper, we use one mass model (FRDM, \cite{Moeller.etal:1995}) and corresponding neutron capture cross sections calculated with the statistical model code NON-SMOKER \cite{Rauscher:2000}. 

\begin{figure}[h]

\begin{center}

\includegraphics[trim=1cm 0.5cm 1.5cm 0cm,clip=true, width=0.6\linewidth,angle=-90]{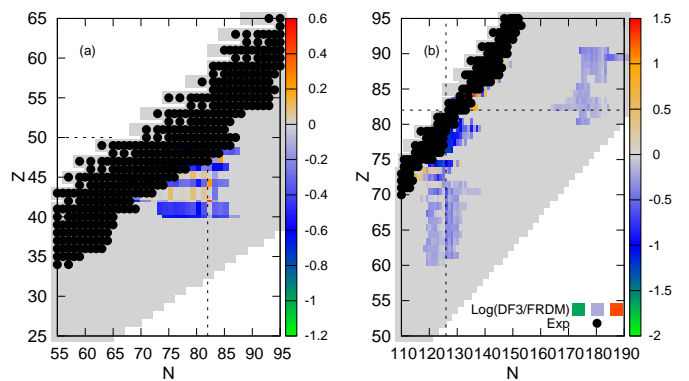}

\caption{Logarithm of the ratio of the two sets of $\beta$ decay rates used in our
 nucleosynthesis calculations. The black points mark the nuclei
  with experimental half lives. The dashed lines correspond to magic numbers.}

\label{fig:rates_exptheo}

\end{center}

\end{figure}

In order to study the effect of beta decays we employ two different sets of half-lives
for those nuclei for which these quantities are not known experimentally. 
The first set, which we take as our reference calculation, is consistent with the FRDM mass model \cite{Moeller.etal:2003}. The second one employs the new rates from Refs. \cite{Borzov:2003,Borzov:2008,Borzov:2011,Borzovnew}. These calculations are based on the FRDM masses, but use energy-density functional theory (a modified version of the functional DF3 derived by Fayans) to calculate the ground state properties and continuum QRPA of finite Fermi System Theory to determine the $\beta$-strength functions. As these self-consistent calculations in general supply good agreement to experimentally known half lives it is expected to exhibit some predictive power for those neutron-rich nuclei on the r-process path with unknown half lives. We point out that these calculations include contributions from Gamow-Teller and first-forbidden transitions. The latter become increasingly important for heavier nuclei and, for example, shorten the half lives of neutron-rich nuclei around $N=126$ significantly. This has recently been confirmed in shell model calculations of r-process waiting point nuclei. Figure~\ref{fig:rates_exptheo} clarifies which beta decay rates we have adopted in our r-process simulations. Black dots mark the nuclei for which experimental half lives are known and we have used these values in all calculations. In the reference calculation we used half lives obtained by QRPA calculations on top of the FRDM model for all other nuclei (called FRDM rates in the following). In the second set (called DF3 in the following) we replace the FRDM half lives by those derived on top of the density functional, which, however, are only available for a limited set of nuclei, including, however, the important r-process waiting points around the magic neutron numbers. For the nuclei, for which Borzov has calculated half lives, Fig. ~\ref{fig:rates_exptheo} shows the logarithm of the ratio between the DF3 (new) and FRDM  (old) rates, where,  for better visualization, the left panel shows only the region around N=82, and the right panel the region around N=126. We observe that the new half lives are in general somewhat shorter. We note that Borzov has not calculated the half lives of some odd-odd nuclei in the $N=82$ region. Hence in the r-process simulations with the new rates we use the FRDM values for these nuclei, which can lead to some artificial odd-even staggering. However, as odd-odd nuclei have noticeably smaller neutron separation energies, $S_n$, than the neighboring even-even or odd-even nuclei, the properties of the odd-odd nuclei have relatively little impact on r-process simulations.

%------------------------------------------------------------------------------------------------------------------------

\subsection{Other r-process scenarios} 
\label{sec:traj} 
The astrophysical site for the r-process remains an open question, but the extreme neutron-rich conditions required point to explosive environments like core-collapse supernovae, neutron star mergers, and accretion disks \cite{arnould.goriely.takahashi:2007}. Parametric studies help to understand the detailed conditions necessary to produce r-process elements as observed in the solar system and old halo star abundances and to explore the impact of the nuclear physics input (see e.g.~\cite{surman.etal:2009, Mumpower.etal:2012}). In our simulations we want to explore the impact of beta decay rates on the r-process dynamics and abundances within the currently favorized astrophysical sites by adopting trajectories of hydrodynamical simulations of neutrino-driven winds and neutron star mergers. 

The neutrino-driven wind from the freshly born protoneutron star in a core-collapse supernova has been suggested as the r-process site in Ref. ~\cite{Woosley.etal:1994}. However, steady-state \cite{Thompson.Burrows.Meyer:2001,Otsuki.Tagoshi.ea:2000} and analytic models (e.g., \cite{Qian.Woosley:1996}) as well as current supernova simulations \cite{arcones.janka.scheck:2007,Arcones.Janka:2011,Huedepohl.etal:2010,Fischer.etal:2010,Roberts.etal:2010} do not yield the extreme conditions needed to produce heavy r-process elements (for a review see \cite{AT13}). For our study we use one trajectory ejected 8~s after the explosion of a 15~$M_\odot$ progenitor (model M15-l2-r1 in ref \cite{arcones.janka.scheck:2007}). The neutron-to-seed ratio in this hydrodynamical simulation is not large enough to produce heavy r-process elements. Therefore, we artificially increase the entropy by reducing the density, as it was done in \cite{Arcones.MartinezPinedo:2011}. An increase by a factor of two in entropy ($S \approx 200 k_\mathrm{B}/\mathrm{nuc}$) is sufficient to produce elements up to the third r-process peak ($A\approx 195$). Although artificial our simulation is expected to  mimic the typical evolution of a high entropy wind which is able to synthesize r-process elements. 

For the neutrino-driven wind model our reference calculation corresponds to the same scaled trajectory as adopted in Ref. \cite{Arcones.MartinezPinedo:2011}. These authors identified the resulting nucleosynthesis as a cold r-process, as the temperature drops very fast and, instead of a $(n,\gamma)-(\gamma,n)$ equilibrium (as in the classical r-process), the r-process nucleosynthesis is dominated by a competition between neutron capture and beta decays \cite{Wanajo:2007}.  As a second neutrino-driven wind scenario we take the hot r-process case presented in \cite{Arcones.MartinezPinedo:2011}, where the $(n,\gamma)-(\gamma,n)$ equilibrium lasts until freeze-out.

Motivated by the problems of the neutrino-driven wind models to produce heavy r-process elements, the nucleosynthesis in neutron star mergers have been recently re-visited as an alternative r-process site, see e.g. \cite{Korobkin.etal:2012,Goriely:2011,Roberts.etal:2011}. This scenario was already suggested by \cite{Lattimer:74} and the first nucleosynthesis studies based on simulations have been reported in \cite{Freiburghaus.Rembges.ea:1999}. Here we use a trajectory from a recent hydrodynamical simulation for which a first report on the nucleosynthesis was presented in \cite{Korobkin.etal:2012}.  The simulation corresponds to the coalescence of a binary system of two neutrons stars with $1.4 M_\odot$ each. %------------------------------------------------------------------------------------------------------------------------ %------------------------------------------------------------------------------------------------------------------------ 
\section{Results} \label{sec:results} %------------------------------------------------------------------------------------------------------------------------ 
\subsection{Cold r-process in neutrino-driven winds} 
\label{sec:cold} 
\begin{figure}

\centering

\includegraphics[width=\linewidth]{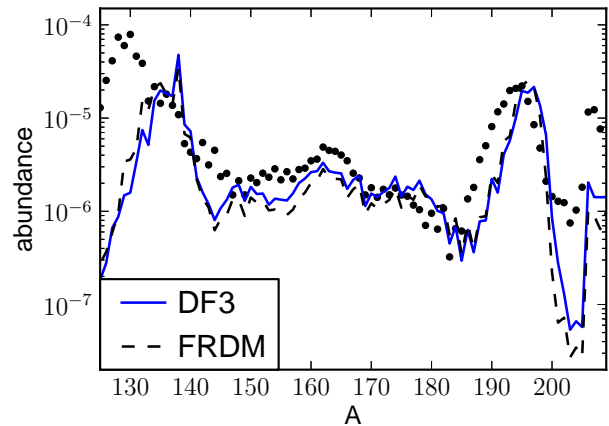}

\caption{Final abundances for cold r-process conditions based on two
sets of beta decay rates (see text for details). The dots represent
  the scale solar abundances.}

\label{fig:ab_allbeta_cold}

\end{figure}

At first we will present our calculations of r-process simulations for neutrino-driven wind conditions which support a cold r-process, i.e. conditions for which the temperature drops very fast and photo-dissociation becomes negligible compared to neutron captures and beta decays (see~\cite{Arcones.MartinezPinedo:2011} for more details). We will discuss this case in some details to illustrate the general effects of the beta decay rates on the r-process dynamics and abundances. Our reference simulation is performed for the FRDM half lives Ref.~\cite{Moeller.etal:2003}. To explore the impact of beta decays on the r-process we repeat the same r-process simulation, however, replacing the FRDM half lives by the DF3 set (Ref.~\cite{Borzov:2008,Borzov:2011}). Figure~\ref{fig:ab_allbeta_cold} shows the final abundances based on the two sets of beta decays. The solar system abundances (dots, \cite{solar}) are included to guide the eye and we do not attempt to reproduce them. Rather our aim is to understand the differences induced by the changes in beta decays.

Our two beta decay sets differ mainly for nuclei in the regions of the magic neutron numbers $N=82$ and $N=126$. Therefore, one expects abundance differences between our two r-process simulations in the region of the second and third peak, i.e., $A\approx130$ and $A\approx195$, respectively (as has been discussed in previous works in the context of the classical r-process, see e.g \cite{Kratz.Bitouzet.ea:1993}). We indeed find these local changes, which, however, are  relatively small (see Figure {\ref{fig:ab_allbeta_cold}}) due to compensating global effects which we explore and explain below. Upon closer inspection of the abundances in Figure {\ref{fig:ab_allbeta_cold}} we find also differences in other regions than the second and third peaks. We can hence conclude that the changes in the beta decay rates have two effects on the r-process abundances: a local which affects the abundances in the region of nuclei with changed half-lives and a global that changes the general abundances pattern, including nuclei which half lives have not been modified. Global changes occur because changes in the nuclear half lives influence the flow of matter to heavier nuclei. Furthermore, such changes affect how fast neutrons are exhausted during the r-process with the important consequence that the freeze-out can occur under different astrophysical conditions. This can have important impact on the final r-process abundances.

\begin{figure}

\centering

\includegraphics[width=\linewidth]{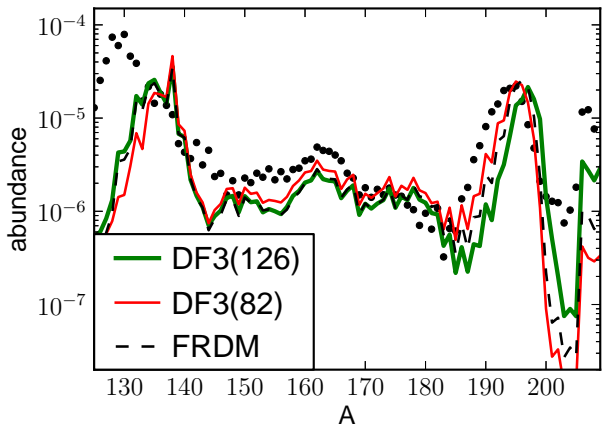}

\caption{Final abundances based on the FRDM set of half-lives and on
  the new rates when used only in the region of $N=82$ (DF3(82)) and only
  around $N=126$ (DF3(126)).}

\label{fig:ab_82_126_cold}

\end{figure}

In order to explore the local and global effects on the final abundance pattern observed in Figure \ref{fig:ab_allbeta_cold} in more details we have performed r-process simulations where we have used the DF3 beta decay rates restricted to the nuclei in the regions around $N=82$ and $N=126$ separately. Figure~\ref{fig:ab_82_126_cold} compares  the final abundances obtained for three different sets of beta decays: 1) our reference calculation with the FRDM rate set (dashed black line), 2) the DF3 beta decays restricted to the region around $N=82$ (called DF3(82), red thin curve) and 3) the DF3 beta decays restricted to the region around  $N=126$ (DF3(126), green thick line).

We observe that using the DF3(82) set lowers the abundances in the second r-process peak around ($A\sim130$). The fact that the DF3 half lives are in general shorter than the FRDM values (see ratios in Figure \ref{fig:rates_exptheo}) leads to a faster flow of matter through this region of nuclei with relatively long half lives. (We mention that the longest half-lives of $N=82$  nuclei on our r-process path are known experimentally (e.g. $^{130}$Cd and  $^{129}$Ag) and are used in all our simulations.) As a consequence a smaller amount of matter is accumulated in this mass region (below $A=138$) when using the DF3 rather than the FRDM half lives. As is argued in  \cite{Arcones.MartinezPinedo:2011} the accumulation of matter at $A=138$, visible by the large abundance peak for all sets of half lives, is a consequence of a problematic behavior of the FRDM masses for $N=82$ which slows down  the r-process matter flow. Nevertheless, this problematic behavior does not change our general conclusion on the faster flow of matter due to the changes in half lives and on changes in the consumption of neutrons which we will discuss later. (We have confirmed this in r-process studies using other mass models. Here we do not want to show these results in order to keep the consistency in mass models used for the nuclear input quantities.) We observe a slightly larger peak at $A=138$ in the DF3(82) simulation. This is a consequence of the faster matter flow through the waiting point nuclei around $A\sim130$ which allows more matter to reach the $A=138$ peak as compared to the FRDM case. 

When using the rate set DF3(126), i.e. using only the DF3 half lives for nuclei around $N=126$,  we observe a local effect in the abundances, now around mass number $A\sim195$, which is similar to the one found for the second abundance peak when using the DF3(82) set. The shorter half-lives predicted by DF3(126) in comparison to the FRDM set lead to a faster flow of matter around ($A\approx 195$) resulting in a slight reduction of the mass accumulated in the peak and to a shift in the peak position towards larger mass numbers.

\begin{figure}

\includegraphics[width=\linewidth]{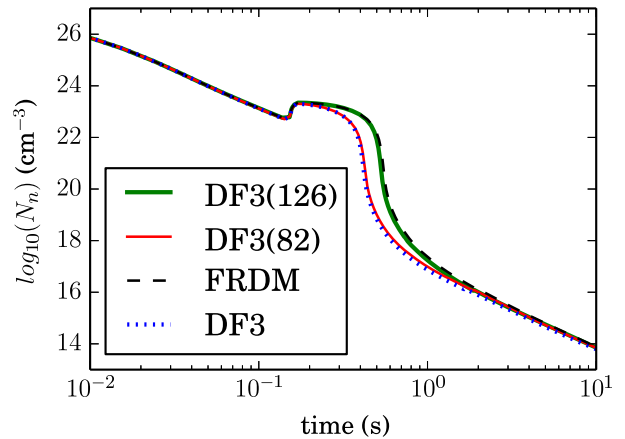}

\caption{Neutron density evolution corresponding to the baseline
  (dashed black line, FRDM) and the calculations with new rates in the
  regions of $N=82$ (DF3(82), thin red line), $N=126$ (DF3(126), thick
  green line), and the full set DF3 (dotted blue line). The small kink at $t\approx 0.2$ is due to the reverse
  shock~\cite{arcones.janka.scheck:2007}.}

\label{fig:nn_82_126_evol}

\end{figure}

So far we have restricted our discussion to local effects; i.e. to changes in abundances in the same regions in which we have changed the half lives. Now we will focus on global effects by examining effects of the half-life modifications on the abundances in other mass regions. The faster mass flow attributed to the shorter DF3 half lives  has two obvious consequences: 1) more matter is moved towards heavier nuclei, this can be seen in Fig.~\ref{fig:ab_82_126_cold} for DF3(82) which has larger abundances for nuclei after the second peak,  and for DF3(126) with larger abundances for nuclei beyond $A=200$, 2) neutrons are consumed faster. This is examplified in Fig.~\ref{fig:nn_82_126_evol} which shows the evolution of the neutron density as function of time. Just after the small kink in the neutron density at $t\approx150$ ms, we observe a noticeable difference in the behavior of the neutron density between the calculations performed for the FRDM and DF3(126) sets on one hand and for the DF3 and DF3(82) sets on the other. For the first two, matter is stalled relatively long around the $N=82$ waiting point, which are overcome faster for the shorter N=82 half-lives used in the DF3 and DF3(82) sets. For the latter case matter moves then faster towards heavier nuclei. This consumes neutrons and hence the neutron density gets reduced. Of course, once the $N=82$ waiting point is overcome for the other two sets of half lives, we find the same reduction in neutron density, however, occuring at slightly later times (and at other conditions in temperature). By closer inspection we find also differences in the neutron density profiles between the calculations which differ only by the N=126 half lives. Using the shorter half lives (as is done with the DF3 and DF3(126) sets) allows for a faster break through the N=126 waiting point and hence also to a faster consumption of neutrons. This is visible in the neutron density profiles which reach the drastic drop at slightly earlier times if we use the shorter N=126 half lives of Borzov in our calculations; i.e. in DF3(126) compared to FRDM and in DF3 compared to DF3(82). These comparisons also indicate that some matter flow starts to pass through the N=126 waiting point already before the strong drop of the neutron density.

\begin{figure}

\includegraphics[width=\linewidth]{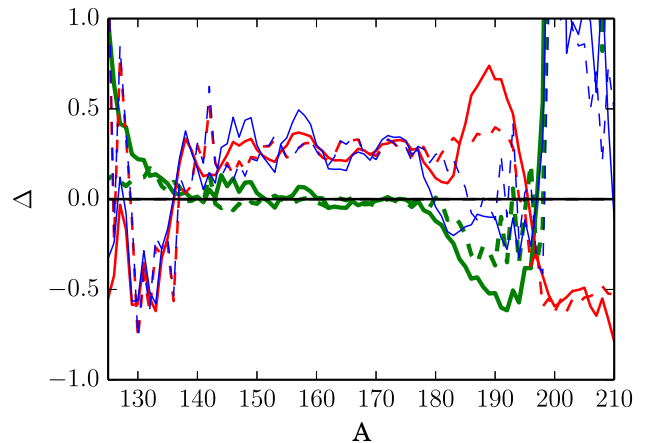}

\caption{Relative differences between baseline abundances (FRDM) and
  abundances based on DF3 (blue thinest lines), DF3(82) (red thin lines) and DF3(126) (green thick
  lines). The solid and dashed lines corresponds to differences in the
  final and freeze-out abundances, respectively.}

\label{fig:delta}

\end{figure}

To visualize the effects of the different half lives on the r-process abundances we define the quantity $\Delta=(Y_{x}-Y_{FRDM})/Y_{FRDM}$ which is the relative difference in abundances obtained when using one of the sets of half lives with the shorter values for N=82 and 126 (i.e. $x$ stands for DF3, DF3(82) and DF3(126), respectively) compared to the benchmark abundances calculated for the FRDM set. The results are shown in Fig.~\ref{fig:delta} for final abundances (solid lines) and at the moment of freeze-out (dashed lines). If we compare the DF3(82) and FRDM calculations, we recover the expected effect: matter is moved from the region around $A=130$ to higher masses due to the faster N=82 halflives. Strikingly this faster matter flow enhances all abundances between the second and third r-process peaks by roughly a constant factor at freeze out. However, we observe less mass flow to nuclei above the A=190 peak for the DF3(82) set than for the FRDM set, which both use the same half lives of these waiting point nuclei. The difference comes from the fact that the neutron density drops much faster for the DF3(82) case implying that there are less neutrons available for capture once the N=126 waiting point nuclei are overcome than for the FRDM case. After freeze-out the relative abundances change slightly. However, there is additional matter accumulated in the A=190 peak for the DF3(82) case. 

If we compare the DF3(126) and FRDM abundances we note relatively little difference in the abundances at freeze out up to the third peak, reflecting the fact that both sets have the same N=82 half lives. However, there is a significant increase for matter beyond A=190, made possible by the faster break through the N=126 waiting point. This is combined with less mass accumulated before the third peak. The enhanced mass transport through the N=126 waiting points brings more matter into the region of heavy nuclei which decay by fission. The fission decay yields of these nuclei produce dominantly nuclei around A=132, which explains the observed relative increase of abundances in the second r-process peak for the DF3(126) calculations, particularly after freeze-out.

Finally we like to discuss the differences between the DF3 and
FRDM calculations. For DF3 the mass flow is faster at the N=82
and 126 waiting points. This leads to interesting differences
with the DF3(82) and DF3(126) cases. For both, DF3 and DF3(82) the
N=82 waiting point is overcome faster than for the FRDM calculations,
leading to a similar abundance pattern upto the third r-process peak.
Furthermore the simulations with DF3 and DF3(82) half-life sets
have similar neutron density profiles
(see Figure \ref{fig:nn_82_126_evol}).
In particular they
reach the nuclei in the region of the third peak with
less neutrons available for captures than in the simulations with the
FRDM half-life set. However, the relatively shorter half lives at N=126
in the DF3 set compared to DF3(82) (and FRDM), allows also for a faster
break through the N=126 waiting point. As a consequence, less matter
is stalled before the third peak and more is transported to heavier nuclei,
where it can also decay by fission contributing to the abundances
in the second peak. Due to the lower neutron density,
and relatively less neutron captures, the simulation with the DF3 half-life set
moves less matter to nuclei beyond the third peak compared to the one using
the DF3(126) set. The combined effect of faster neutron consumption
on one hand (due to the shorter N=82 halflives) and faster flow through
the N=126 region on the other hand result in the interesting feature
that the third abundance peak is shifted slightly towards smaller mass
numbers in the DF3 case compared to the DF3(126) one, resembling the
behavior found for the FRDM half lives.

We like to finish the discussion of our cold r-process scenario with three general remarks. The abundance peak observed in all our calculations at A=138 is probably an artifact of the FRDM mass model. Furthermore, in the calculations using the faster N=126 half lives there is a local flow of material to heavier nuclei which is compensated by an increased flow from $A\sim130$. The net effect is an almost unchanged peak shape. Our calculations show the importance of neutron captures during decay to stability. We conclude that the neglect of neutron captures after freeze-out (i.e. for $Y_n/Y_{seed}=1$), as done in classical r-process simulations, is questionable. As has been discussed previously (e.g.\cite{Mumpower.etal:2012,surman.etal:2009,Surman.Engel.ea:1997}), it is crucial to go beyond the classical r-process picture if one aims at studying the effect of new half lives on r-process abundances. Finally we note that the evolution of the neutron density may also change due to improved beta-delayed neutron emission probabilities $P_n$ (see e.g.\cite{Arcones.MartinezPinedo:2011,Domingo-Pardo:2013}).

%------------------------------------------------------------------------------------------------------------------------ 
\subsection{Local and global effects for different astrophysical conditions} \label{sec:diffastro} 
The change of half lives at the N=82 and 126 waiting points from the FRDM set to the faster DF3 values has similar general effects on the r-process evolution in the two other r-process scenarios which we have studied (hot r-process in the neutrino driven wind and neutron star mergers) as we have discussed above for the cold r-process: The shorter DF3 half lives allow the matter flow to overcome the waiting points faster and more matter is transported to nuclei between the two r-process peaks and in particular to heavy nuclei beyond the third peak at $A \sim 195$.  However, there are features specific to these two r-process sites on which we want to focus in the following. To illustrate our results in more details we have again performed calculations in which we have only replaced the FRDM halflives by the DF3 values in the neighborhood of the N=82 waiting point (again denoted by DF3(82)) and at N=126 (DF3(126)). 

\begin{figure}

 \includegraphics[width=\linewidth]{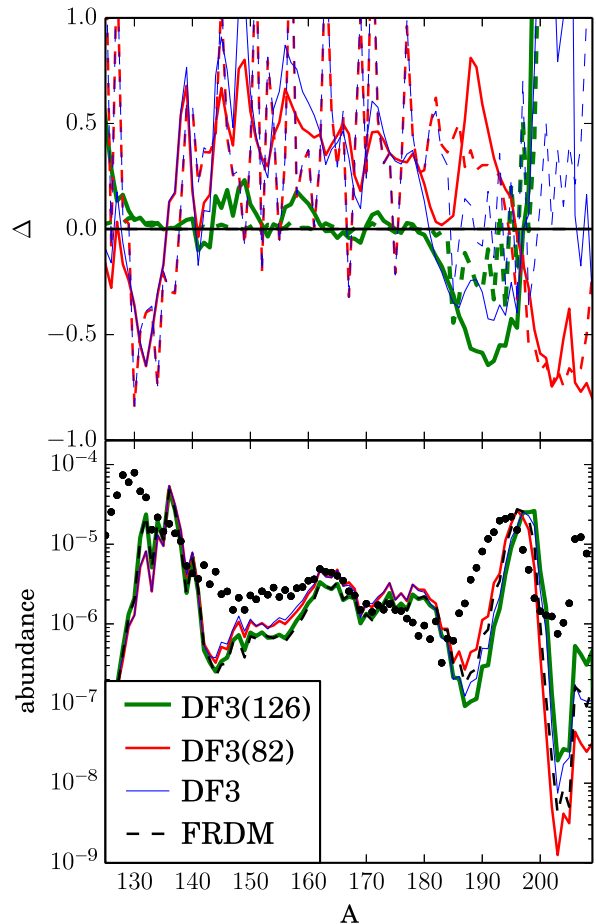}

  \caption{Abundances based on different sets of beta decay rates
    (same as in Fig.~\ref{fig:ab_82_126_cold}) for a neutrino-driven
    trajectory characterized by a hot r-process.}

  \label{fig:ab_delta_hot}

\end{figure}

One of the important features of the hot r rprocess in the neutrino driven wind is the fact that the reverse shock keeps the temperature relative high ($T\approx 1$~GK) for some time (see \cite{Arcones.MartinezPinedo:2011} for more details). Under these conditions an $(n,\gamma)-(\gamma,n)$ equilibrium can be achieved and the evolution, before freeze-out, is similar to the classical r-process. The equilibrium implies that, for given temperature and neutron density, the r-process path is defined by the neutron separation energy of isotopes involved. Figure~\ref{fig:ab_delta_hot} shows the final abundances of r-process simulation for the various sets of half lives and their relative differences, indicated by the quantity $\Delta$ as defined above. We observe the same general trends as discussed above due to the faster mass flow. However, upon closer inspection we observe several interesting details. In all calculations (DF3, DF3(82), DF3(126)) there is less total abundance accumulated in and, in particular, beyond the third r-process peak in the hot scenario than found in the cold r-process case (compare to Figure \ref{fig:delta}). This observation is consistent with the fact that the r-process path in the hot scenario runs, due to the higher temperature, through nuclei closer to stability. These have generally somewhat longer half lives than the more neutron-rich nuclei encountered on the path of the cold r-process, in particular at the waiting points. These longer half lives slow down the flow of matter to heavier nuclei and hence less matter is transported into the region of the nuclear chart where nuclei decay by fission. A striking feature of the $\Delta$ plot are the singular changes in abundances for selected nuclei at freeze-out observed in the DF3(82) calculation. This is caused by the fact that the faster break through the N=82 waiting point in the DF3(82) (and DF3) calculation, compared to the FRDM case, leads to a faster consumption of neutrons, as discussed above for the cold r process. As a consequence, the $(n,\gamma) - (\gamma,n)$ equilibrium, which is valid before freeze out, favors another neutron separation energy and as a consequence changing the abundances within isotopic chains. As is wellknown such effects are washed out after freeze out when matter decays to stability, which is also observed in our calculations.

Due to the extremely neutron-rich environment the r-process in neutron star mergers runs through nuclei close to the neutron dripline. In general this means that relatively short half lives are encountered on the path and much material is transfered into regions of heavy fissioning nuclei. In fact, we observe fission cycling in all our calculations where we have used a trajectory from a recent hydrodynamical simulation (Sect.~\ref{sec:traj} and \cite{mergertraj}). This means that independently of the set of half lives used matter is encountering several transport cycles from medium mass nuclei to heavy fissioning nuclei, which by fission produce medium-mass nuclei. Under these conditions the abundances in the second r-process peak are produced mainly as fission yields. 

\begin{figure}

\includegraphics[width=\linewidth]{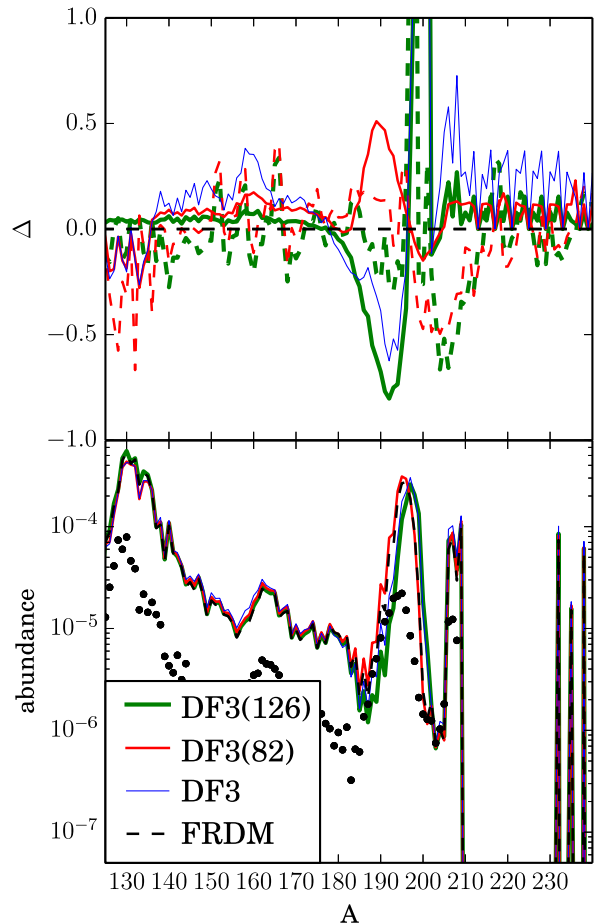}

  \caption{Same as Fig.~\ref{fig:ab_delta_hot} for a trajectory of a
    neutron star merger simulation.}

  \label{fig:ab_delta_merger}

\end{figure}

The fact that the r-process path moves through nuclei close to the neutron dripline has the 
following consequences for our r-process simulations with different half lives. 
For the N=82 waiting point our two sets of half lives (FRDM, DF3) are quite similar 
for nuclei close to the dripline. 
Hence these waiting point nuclei are overcome fast and in the same dynamical time scale in 
all calculations. Thus we do not observe more matter between the second and third peaks when 
comparing the DF3(82) and FRDM calculations (see Figure~\ref{fig:ab_delta_merger}). 
This is in contrast to the cold and hot neutrino-driven wind models. At N=126 the DF3 half lives 
for neutron-rich nuclei are shorter than the FRDM ones. These differences, however, are more 
pronounced when moving closer to stability. (We note that one important reason for the 
shorter half lives is the incorporation of forbidden contributions.) The differences 
in half lives have the following dynamical effect if we compare the  DF3 and DF3(126) 
calculations with the FRDM ones. Once the matter flow reaches the heavy nuclei beyond 
the N=126 waiting point, these nuclei fission and their decay yields are distributed back to 
the second r-process peak region. As the r-process proceeds, the neutron density decreases 
and the path moves closer to stability reaching a region of nuclei 
for which the FRDM half lives are longer than the DF3 values 
Then the shorter half-lives in the DF3 simulations allow for a faster transfer  
of material from the second peak region to the third peak region and beyond. 
In summary, the main effect of the shorter half lives in the set of DF3 calculations is 
a change of time scale for the fission cycles; i.e. matter moves faster from lighter nuclei 
to heavier (as is the feature in the DF3(82) calculation) and can be transported faster 
beyond  the third peak where it eventually will decay back by fission producing new 
material around the second peak. We stress that the fissioning nuclei in our calculations 
are located around mass numbers $A=280$. Hence the fission yields of these nuclei do not 
directly populate the third r-process peak. Hence interesting local abundance features, 
even after freeze out, can occur in 
the mass region of the second peak, which we will discuss briefly in the following.  
In particular, one observes in Figure~\ref{fig:ab_delta_merger} that the noticeable underproduction of abundance in the DF3 calculations around the second peak observed at freeze-out is filled up by yields from late fission. In fact we find that, independently of the set of half lives used, the final r-process abundance distributions look quite similar up to masses around the third peak. Here the differences introduced by the faster half lives do not only pertain, they are modified and partially enhanced after freeze out. At first, some of the late fission yields, originally produced as matter around the second r-process peak, is transported towards the third peak by neutron captures, even after freeze out. With the faster DF3 half lives more matter can even at late times overcome the N=126 waiting point than for the FRDM half lives, explaining the growth of the abundance trough around A=195 in the relative abundance differences $\Delta$. It is also interesting to note that $\Delta$ is increased for a few nuclei around $A=200$, related to the longest half lives encountered at the third peak. At first, long half lives, which act as obstacles on the r-process path, are already encountered for smaller proton numbers at N=126 for the longer FRDM half lives. As one consequence, matter is accumulated at slightly lower mass numbers when using the FRDM N=126 half lives. Secondly, we notice that the nuclei, which are strongly produced when using the DF3 halflives, are moved up in mass number slightly after freeze out. This is due to the fact that the r-process path moves closer to stability due to decreasing neutron density.

Finally we note that after freeze out matter between the third peak and the long-lived 
thorium and uranium isotopes decays either by fission or $\alpha$ decay, explaining the abundance trough
in the final r-process abundance beyond the third peak. 

%------------------------------------------------------------------------------------------------------------------------

\section{Conclusions}

\label{sec:conclusions}

Due to progress in nuclear modelling and constrained by half-life measurements
at RIB facilities for nuclei at or towards the r-process path, indications
have grown that the half lives of key r-process nuclei close to the magic
neutron numbers are shorter than anticipated in global models usually
used in r-process simulations. In particular evidence has grown that 
forbidden transitions, until recently not being considered adequately
in half-life calculations, contribute significantly to the beta decay
rate for nuclei in the vicinity of the N=126 waiting point. We have studied
the impact of the potentially shorter half lives on the dynamics and the
abundances of the r-process. To do so, we have performed calculations
for two different sets of half lives: in the first we have adopted
the predictions of the global FRDM model for those nuclei for
which experimental values are not available, in the second set we have
replaced the FRDM half lives by those of the DF3 model. The latter model 
gives a fair account also of the recent experimental data and generally
predicts faster half lives than the FRDM model, making it an appropriate
tool to study the effect of shorter half lives on r-process simulations. 
However, the DF3 model
is restricted to a selected set of spherical nuclei, which importantly
includes the nuclei in the region around the N=50, 82 and 126 r-process
waiting points. Our r-process simulations have been performed for three
different astrophysical scenarios: a cold and hot r-process related
to the neutrino-driven wind scenario in supernova explosions, and
an r-process in the extremely neutron-rich environment encountered in
neutron star mergers. To study the impact of shorter half lives on the 
r-process in more details we have also performed calculations in which
we only partially use the DF3 half lives; i.e. in which the substitution
of the FRDM half lives by the DF3 ones is restricted to the region
of either the second (N=82) or the third (N=126) r-process peak.

The general feature of the calculations with the shorter half lives (DF3 set)
is a faster mass flow of
matter towards heavier nuclei as the relatively long waiting point nuclei
at N=82 and 126 are overcome faster. Associatedly
neutrons are consumed faster and the r-process freeze-out occurs
at different astrophysical conditions. These effects on the r-process dynamics
have interesting consequences for the final r-process abundances. In the cold
r-process scenario the faster mass flow through the N=82 waiting point moves
more matter beyond A=130. However, the mass flow through the N=126
waiting point and beyond is governed by two competing effects:  
the shorter half lives support a faster flow through the N=126
waiting point, however, there are less neutrons 
available when the matter flow reaches these nuclei throttling 
the production of heavier nuclei beyond A=195. The result
of these competing effects is that
the final r-process abundance in the third peak is quite similar 
in simulations with the faster (DF3) and longer (FRDM) half-life sets.
We note that this is different if we only replace the half lives in the
region of the N=126 nuclei, as then the break through the N=126 waiting point
occurs at a significantly higher neutron density.

For the hot r-process scenario we find quite similar trends when
comparing the calculated abundances for the shorter and longer half lives.
However, there is less matter produced beyond the third (A=195) r-process peak
in the hot scenario caused by the fact that, due to the higher temperatures
involved in this astrophysically site, the r-process path runs through
nuclei closer to stability than in the cold scenario. 
These nuclei have longer half lives and serve as stronger obstacles,
particularly at the N=82 and 126 waiting points, for mass flow towards
heavy nuclei. We also observe that the faster break through the N=82
waiting point nuclei, associated with the faster consumption of neutrons,
can change the conditions of the ($n,\gamma)-(\gamma,n)$ equilibrium,
which is achieved in the hot r-process scenario before freeze out, 
and might favor other nuclei in an isotopic chain. This leads to significant local changes in abundances during the r-process, but
is mostly washed out during decay to stability.

The r-process mass flow in the neutron star merger scenario is characterized
by the fast half lives (compared to the neutrino-driven wind model) of
the nuclei on the r-process path close to the neutron dripline, 
transporting much material to heavy nuclei into the fissioning 
region beyond the thrird peak
and supporting several fission cycles. The main result of the faster
half lives in the DF3 set is a reduction of the fission cycle time allowing
for a faster transport of matter from the second r-process peak region (to
which the fissioning nuclei in our simulations with mass numbers around
$A \sim 280$ dominantly decay) to the region of fissioning nuclei.
We find that the matter flow from the fission yields towards heavier
nuclei can continue even after freeze out and that at late times shorter
half lives in the N=126 waiting point region can induce quite significant
abundance changes in this region.

In summary, our r-process studies have highlighted some effects of the
half lives of waiting point nuclei in the N=82 and 126 mass regions
on the final abundances. This obviously calls for more experimental data,
either to determine some of the key half lives directly or to constrain
theoretical models and hence to reduce the uncertainties in the predictions.
Such important experimental progress is expected from radioactive
ion-beam facilities like RIKEN or in the future at FAIR and FRIB, which
promise to reach also the important waiting point nuclei at the third
r-process peak.

\begin{acknowledgments} The authors thank C. Domingo-Pardo, F. Montes, and R. Surman for helpful and stimulating discussions. This work was supported in part by the Helmholtz Alliance Program of the Helmholtz Association, contract HA216/EMMI ``Extremes of Density and Temperature: Cosmic Matter in the Laboratory'' and the Helmholtz Nuclear Astrophysics Virtual Institute (NAVI, VH-VI-417). A.A. is supported by the Helmholtz-University Young Investigator grant No. VH- NG-825. G.M.P. acknowledges support by the Deutsche Forschungsgemeinschaft through contract SFB 634 and the Helmholtz International Center for FAIR within the framework of the LOEWE program launched by the state of Hesse. I.B was supported by Helmholtz Alliance Program of the Helmholtz Association EMMI during his visit to GSI and by IN2P3-RFBR agreement 110291054. \end{acknowledgments}

%------------------------------------------------------------------------------------------------------------------------

%\bibliography{biblio_beta}

\end{document}